\documentclass{article}
\usepackage{arxiv}

\AtBeginDocument{%
  }

\usepackage{multirow}
\usepackage{tabu}
\usepackage{url}
\usepackage{verbatim} 

\usepackage[utf8]{inputenc} 
\usepackage[T1]{fontenc}    
\usepackage{hyperref}       
\usepackage{url}            
\usepackage{booktabs}       
\usepackage{amsfonts}       
\usepackage{nicefrac}       
\usepackage{microtype}      
\usepackage{lipsum}		
\usepackage{graphicx}
\usepackage{subfig} 

\begin{document}
\title{D4R - Exploring and Querying Relational Graphs Using Natural Language and Large Language Models - the Case of Historical Documents}


\author{
  Michel Boeglin \\
  IRIEC -- Univ. Paul-Valéry \\
  Montpellier, France \\
  \texttt{michel.boeglin@univ-valery.fr} \\
  \And
  David Kahn \\
  INU Jean-François Champollion \\
  FRAMESPA, UMR5136 CNRS \\
  Albi, France \\
  \texttt{david.kahn@univ-jfc.fr} \\
  \And
  Josiane Mothe \\
  INSPE, UT2J, IRIT, UMR5505 CNRS \\
  Univ. de Toulouse \\
  Toulouse, France \\
  \texttt{josiane.mothe@irit.fr} \\
  \And
  Diego Ortiz \\
  IRIT, UMR5505 CNRS, Univ. de Toulouse \\
  Toulouse, France \\
  \texttt{diego.ortiz@irit.fr} \\
  \And
  David Panzoli \\
  IRIT, UMR5505 CNRS \\
  INU Jean-François Champollion \\
  Albi, France \\
  \texttt{david.panzoli@univ-jfc.fr} \\
}



\maketitle

\begin{abstract}
D4R is a digital platform designed to assist non-technical users, particularly  historians, in exploring textual documents through advanced graphical tools for text analysis and knowledge extraction. By leveraging a large language model, D4R translates natural language questions into Cypher queries, enabling the retrieval of data from a Neo4J database. A user-friendly graphical interface allows for intuitive interaction, enabling users to navigate and analyse complex relational data extracted from unstructured textual documents. Originally designed to bridge the gap between AI technologies and historical research, D4R's capabilities extend to various other domains. A demonstration video and a live software demo are available.
\end{abstract}


\section{Context and Objectives}
\label{sec:Introduction}

Historians and researchers frequently analyse large volumes of textual documents to extract meaningful insights about historical events, people, and relationships. They try to answer questions such as ``Who were the key figures in a specific place or time?'' or ``What are the links of a family or group of people?''. However, manually processing these vast collections is time-consuming and error-prone. While text mining and natural language processing (NLP) tools exist, they often require technical expertise, making them inaccessible to many historians.

To address these challenges, we introduce D4R (D4R for Religious Dissent and Reception of the Reformation during the Renaissance, Spain – 16th century), a digital platform designed to enable historians to explore and query historical texts using natural language. D4R allows users to refine their research questions incrementally through a user-friendly graphical interface. D4R leverages a large language model (LLM) to translate natural language queries into Cypher queries, which retrieve data from a Neo4J graph database. This combination of advanced technology and intuitive design makes D4R a powerful tool for historians to work with large amounts of historical text and uncover meaningful insights.

D4R was initially developed to support historical research, but its capabilities extend to various domains that require graph-based data exploration. In this paper, we demonstrate D4R’s application in analysing historical trial records, showcasing how relational graphs can be effectively explored using LLM-assisted query generation.

A demonstration video is available at ~\url{https://www.youtube.com/watch?v=mxoS4QFwHog}, and attendees will have the opportunity to engage with the platform during the conference (~\url{https://d4r-app.irit.fr/sigirdemo/}). 

\section{System Description}
\subsection{Architecture}

\begin{figure}[h]
    \centering
    \subfloat[Node types distribution]{
        \includegraphics[width=0.49\linewidth]{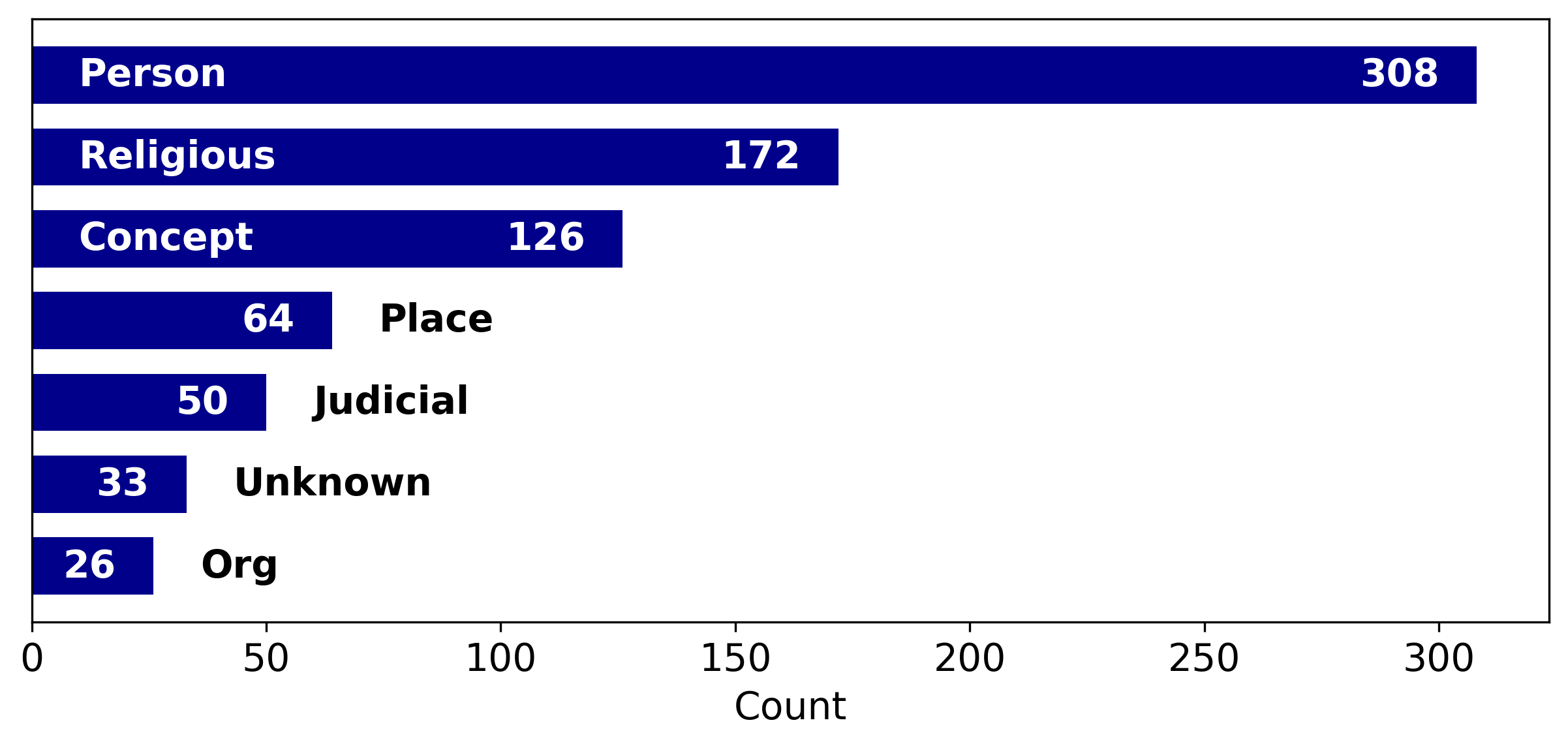}
    }
    \hfill
    \subfloat[Relationship types distribution]{
        \includegraphics[width=0.48\linewidth]{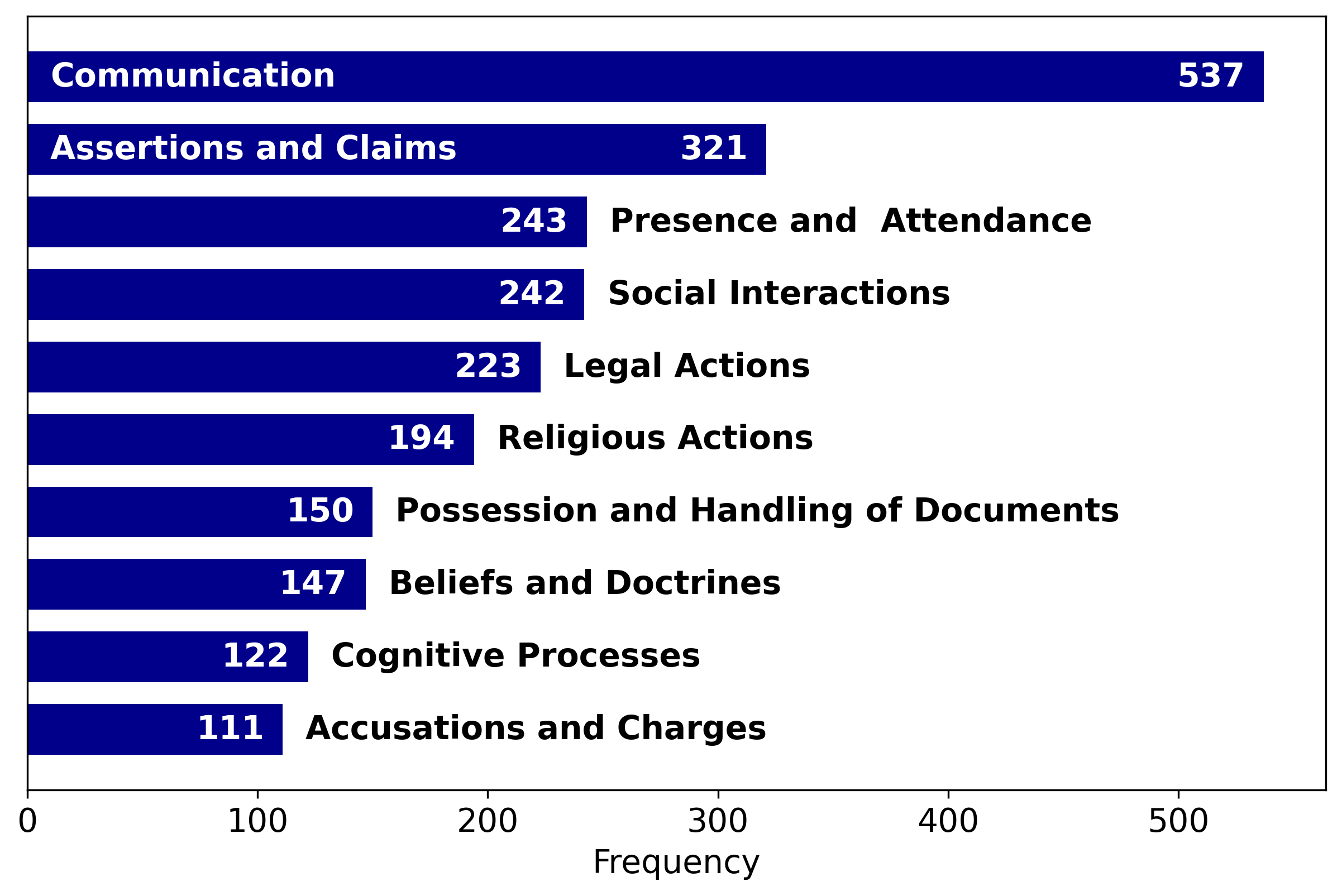}
    }
    \caption{Left: Distribution of node types in the historical dataset by category. Right: Distribution of relationship types in the historical dataset.}
    \label{fig:distributions}
\end{figure}

D4R is based on relational data, which includes named entities (like people and places) and the connections between them. This structure keeps the data both detailed and adaptable, allowing users to easily explore and create queries for various purposes.
The graph schema features different types of named entities, represented as nodes. These can be general categories such as Person, Place, and Organisation, or domain-specific categories like Religious and Judicial entities.
Additionally, a paragraph node stores each paragraph's text along with its metadata, such as the paragraph ID, type, archival source details, and folio or page numbers.
The relationships between these entities are extracted from the text and categorized into 10 predefined types (listed in Figure~\ref{fig:distributions}) which were validated by domain experts, ensuring their relevance and accuracy for historical analysis.


Each node and relationship in the database is enriched with attributes that provide contextual information:
\begin{itemize}
    \item \textbf{Nodes}: Each entity node (excluding paragraph nodes) includes a label (e.g., 'Person', 'Place') and metadata such as the text of the paragraph from which it was extracted. 
    \item \textbf{Relationships}: Each relationship between two entities is associated with a relationship type, its category, and the sentence in which it was extracted. This ensures that historians can trace the information back to its original source.
\end{itemize}

Details of the graph extraction process are beyond the scope of this paper and are discussed in our ECIR 2025 paper on relation extraction~\cite{hidalgo2024adapting}. D4R provides visualization tools for relationships and nodes, enabling users to explore the data and better understand entity connections.

\subsection{Dataset}

The original dataset used in the D4R platform is a private historical dataset focused on the trial of Pedro de Cazalla during the Spanish Inquisition. It consists of approximately:
\begin{itemize}
    \item \textbf{600 nodes}, representing named entities such as people, places, and organisations.
    \item \textbf{3,000 relationships}, representing the connections and interactions between entities.
    \item \textbf{13,000 properties}, associated with both nodes and relationships, including detailed text and metadata from the trial documents.
\end{itemize}
Although this dataset is not publicly available, it serves as the primary resource for testing and refining the D4R platform in a historical research setting. 

The extraction of named entities and relationships from these documents allows for an assessment of the most frequently represented categories. As shown in Figure~\ref{fig:distributions} (top), the majority of named entities are people, with a particular emphasis on religious figures. Regarding  relationships (see Figure~\ref{fig:distributions}, bottom), the two most prominent categories confirm that the dataset primarily  involves verbal exchanges and claims, like accusations and legal arguments. The data distribution aligns well with the research needs of historians, as described in the use case in section~\ref{sec:usecase}.

\subsection{Flexibility for Other Domains}
The D4R schema is designed to be flexible, allowing it  to handle data from various domains. Since its primary focus is on relationships between named entities, datasets from other fields, such as news articles or corporate data, can be easily  imported and processed. 

As an example, D4R was successfully tested using the CoNLL04 dataset \cite{carreras-marquez-2004-introduction}, a benchmark dataset for relation extraction. This dataset consists of 1,437 sentences from news articles, annotated with entity categories such as 'person', 'organization',  'location', and 'other'. The dataset also includes relationships types like 'kill', 'work for', 'Organization-based-in', 'Live in', and 'located in'. This showcases D4R's adaptability,  as it can efficiently process different types of data while maintaining both data integrity and usability. The CoNLL04 dataset was preprocessed to ensure compatibility with D4R. This involved adapting  entity types and relationships to match the format required by the system, as well as ensuring that the dataset’s text and relationships could be properly visualized within the D4R interface.

Although the private historical dataset and CoNLL04 differ in content (historical trial data vs. news articles), both share a relational structure. D4R does not require explicit entity mapping and can efficiently handle datasets containing entities and relationships, making it highly versatile  across various domains. This shows D4R can support diverse datasets while providing meaningful visualizations and query capabilities,  regardless of the application domain.

\section{Querying Workflow: From Natural Language Queries to Cypher Queries }
\begin{figure}[h]
\centering
\includegraphics[width=\linewidth]{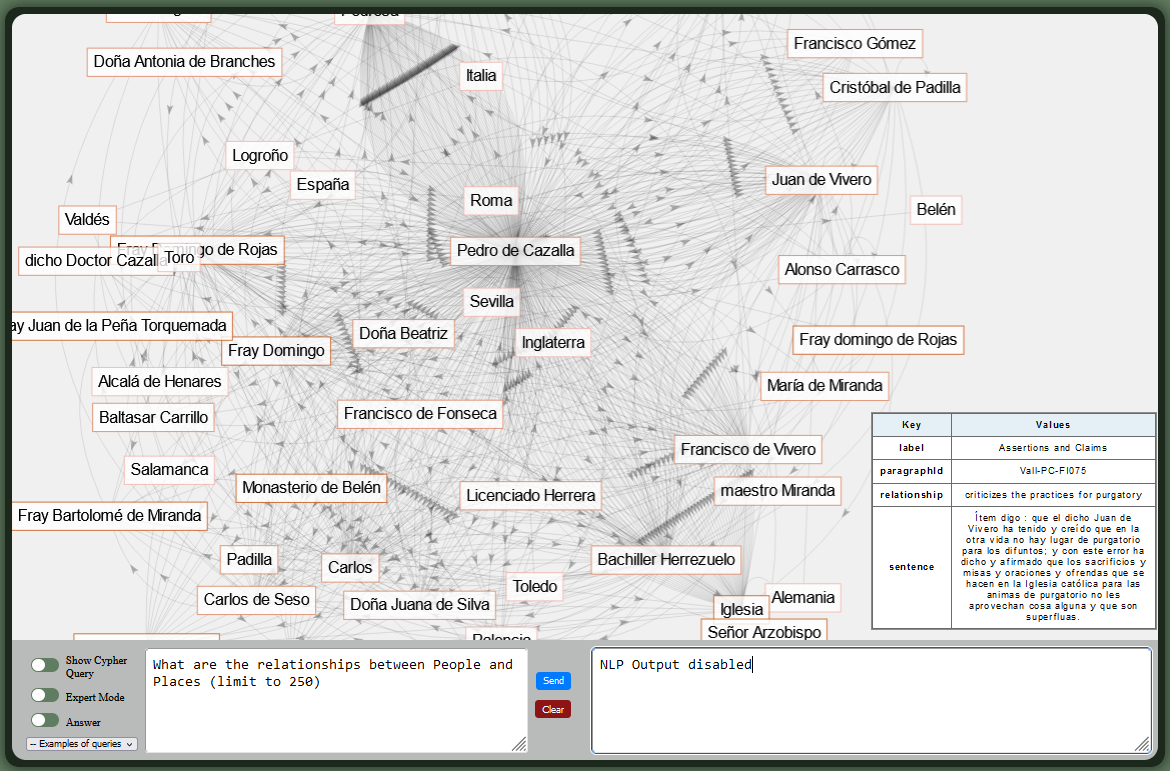}
\caption{D4R presents any text corpus as a graph, where entities and relationships were automatically extracted. Users can query this graph by formulating  natural language questions, which are then automatically translated into Cypher queries to retrieve information from the Neo4J database.}
\label{fig:d4r}
\end{figure}

In D4R, content is represented as a relational graph, as shown in Figure~\ref{fig:d4r}. Named entities extracted from the texts are displayed graphically, along with their relationships.

D4R stores this graph-based content in a Neo4j graph database~\cite{guia_graph_2017}, using Cypher~\cite{francis_cypher_2018} as querying language. To enable non-technical users  to interact with the graph structure,  natural language  questions must be translated into Cypher queries. A similar approach has been successfully implemented for SQL relational language~\cite{hong2024}.
Here, this translation  is performed using a large language model (LLM). We select GPT-4o mini \cite{openai_gpt-4_2024} for its cost-effectiveness and strong performance, achieving  satisfactory accuracy in generating graph queries while minimizing token expenditure. Other alternatives could have been evaluated.

\begin{figure}
\centering 
\includegraphics[width=\linewidth]{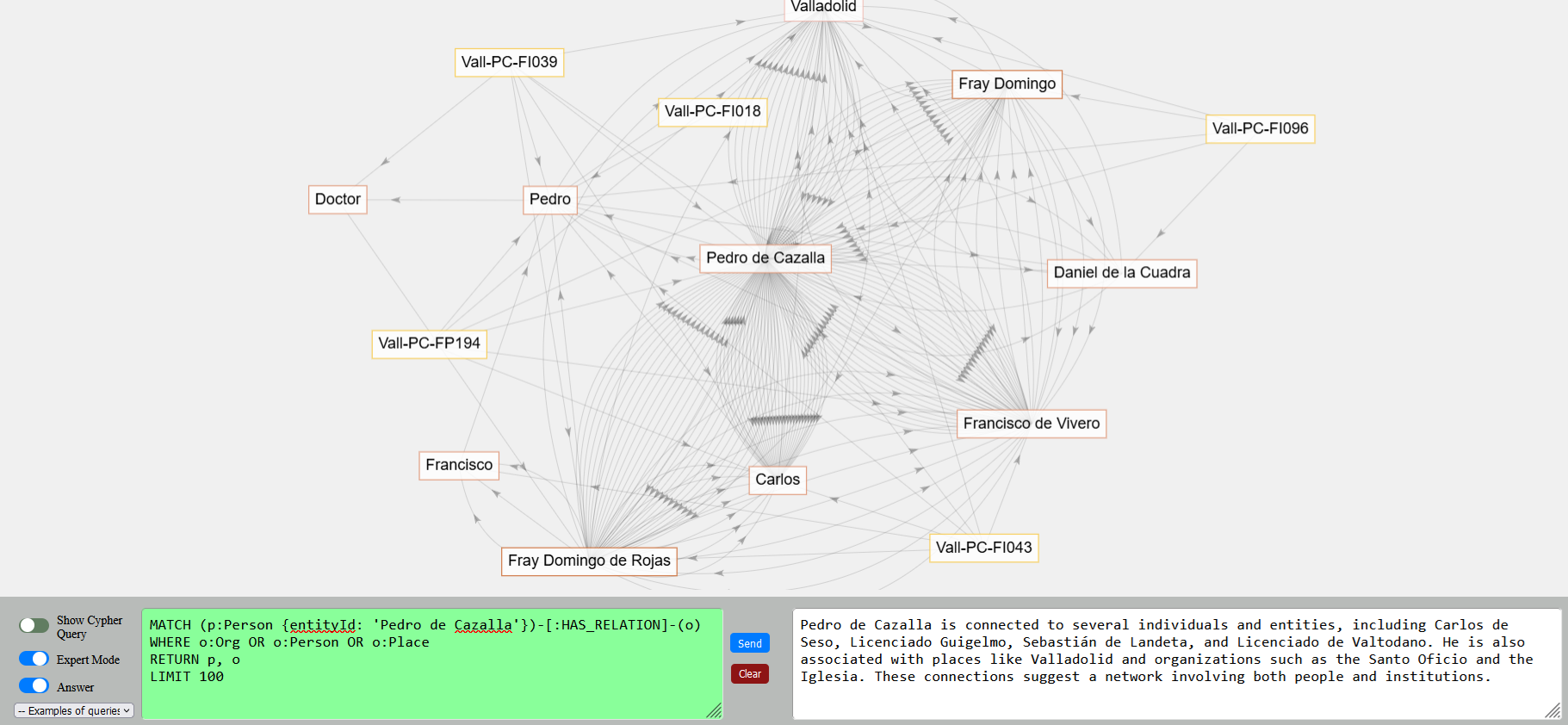} 
(a)
\includegraphics[width=\linewidth]{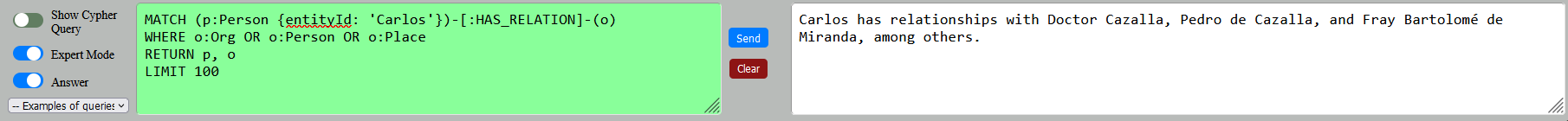} 
(b)
\caption{(a) A user query, expressed in natural language, is translated into Cypher syntax for execution in the Neo4j database. The result is a subgraph,  displayed in the main window, along with a natural language response. (b) D4R's expert mode enables users to edit Cypher queries directly, provided they have the necessary expertise.} 
\label{fig:cypher}
\end{figure}

The first step in processing a query is to map the named entities on the graph. A specialized prompt, tailored for named entity recognition, guides the LLM to extract relevant entities--~such as people, organisations, locations, or paragraph IDs~--from the user's query. 
The extracted entities are then mapped to their corresponding representations in the Neo4j database. This mapping utilizes a fuzzy matching technique to account for variations in entity names, ensuring that the extracted terms align with the database records.
%
The second step consists in crafting a ``schema-aware'' Cypher query. The LLM is used again to construct the query, informed by Neo4j's graph schema, which details node types and relationships. Including schema information in the prompt is essential for generating effective queries. This ensures that the queries accurately align  with the data's structure and relationships, leading to reliable and meaningful answers.
Additionally, the prompt is designed to enforce  the two following rules:
\begin{itemize}
    \item \textbf{Bidirectional Relationship Matching}: 
The model checks relationships in both directions between entities. In  historical research, this  is crucial for understanding both incoming and outgoing relationships.
\item \textbf{ Only direct relationships} between any two nodes are considered, excluding indirect paths obtained through transitive relationships, as they are  unnecessary given the graph schema and would significantly increase processing time.
\end{itemize}
%

Finally, the Cypher query is executed on the Neo4j database, and the corresponding results are retrieved. The prompt can also be instructed to generate a natural language reponse, aligned with  the original question (see Figure~\ref{fig:d4r}).

\section{User Experience and Visual Interface}

With its automated Cypher query translation, D4R enables users to query a corpus naturally and efficiently. The platform accommodates both non-expert and expert users, offering two usage modes based on the user's familiarity with D4R and Cypher. 
In the basic mode, users formulate questions in natural language (see Figure~\ref{fig:cypher}.a).
For instance, a historian may input a query to identify connections between key historian figures or  in the historical corpus studied, or their pertaining relationships with organisations, places, or religious concepts. 
To achieve this, the user simply inputs their query into the designated text field. The system translates and executes the query, yielding two results. 
First, the resulting subgraph is displayed in the main application main window. The subgraph is interactive, allowing the user to zoom in and out, to give focus and  manipulate entities (represented as labels) and their relationships (represented as edges). 
A contextual menu provides quick access to detailed information on entities and relationships with a simple click (see Figure~\ref{fig:d4r}, right). This menu also allows users to trace  data back to its original corpus sources,  ensuring that each entity and relationship is anchored in historical text. By linking  graph data to its original textual source, users can construct their own representations and make informed  interpretations.
The system also generates a natural language  summary of the query results, making it easier to interpret  complex graph data.
This feature simplifies data analysis, ensuring users can  understand relationships and connections without having to manually analyse the visual graph structure in detail.
%
Users also have the option to view the Cypher translation of their natural language query  (see Figure~\ref{fig:cypher}.a). This feature helps users understand  Cypher syntax, gradually building their expertise and allowing them to bypass limitations by querying the platform directly in Cypher.
This simple feature paves the way for more advanced usage, including expert mode (see Figure~\ref{fig:cypher}.b) where  users submit Cypher queries directly—either written from scratch or by modifying previous queries. Apart from bypassing automatic translation, the platform functions in the same way as  standard mode.
%
The following section illustrates a use case where a historian leverages the D4R platform's features to investigate a research question.

\section{Example Use Case}
\label{sec:functionality}\label{sec:usecase}

\begin{figure}
\centering 
\includegraphics[width=\linewidth]{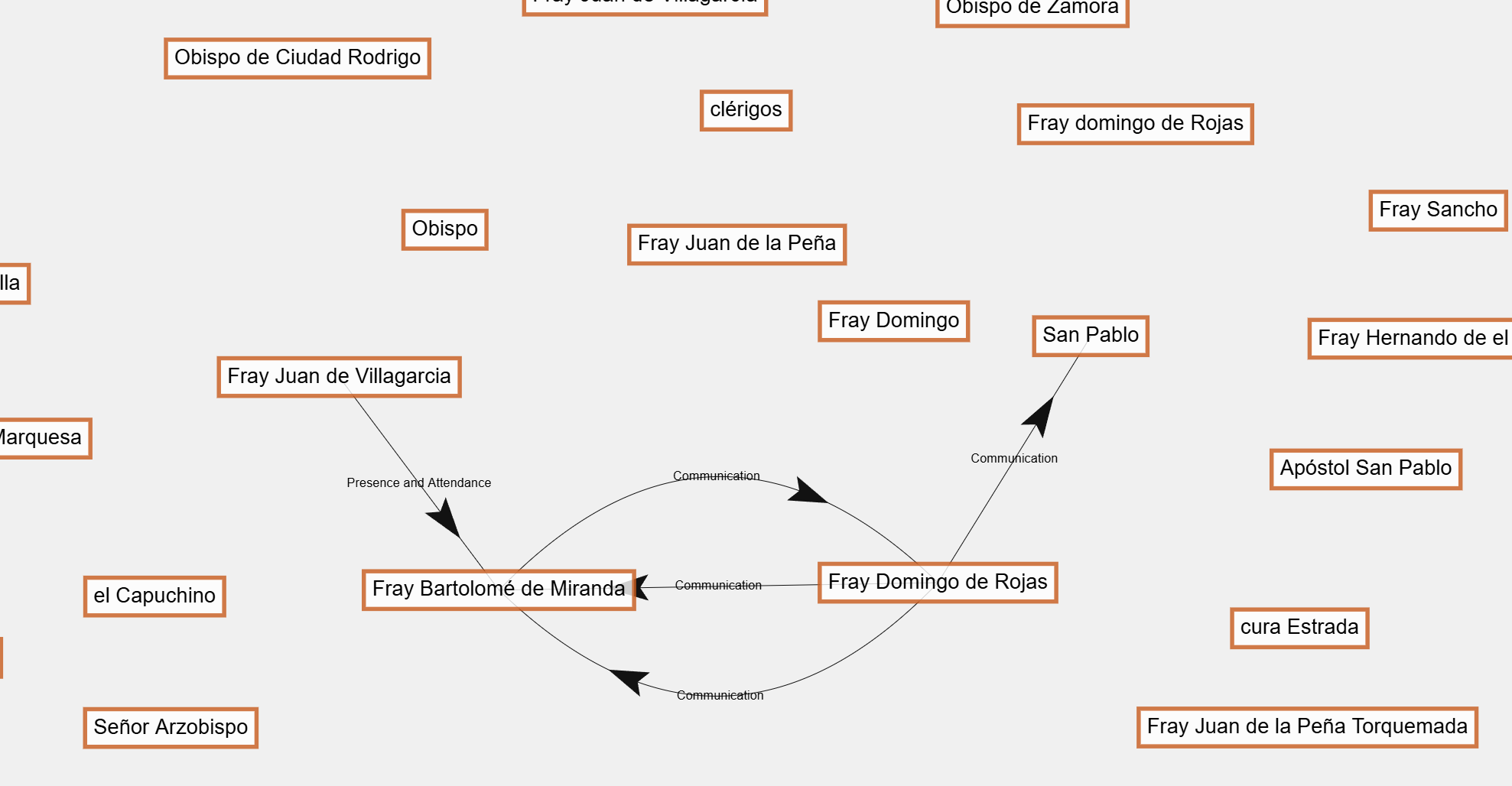} 
\caption{The first query aims to identify entities in the corpus that have both the  “person” and “religious” attributes.} 
\label{fig:query1}
\end{figure}

\begin{figure}
\centering 
\includegraphics[width=\linewidth]{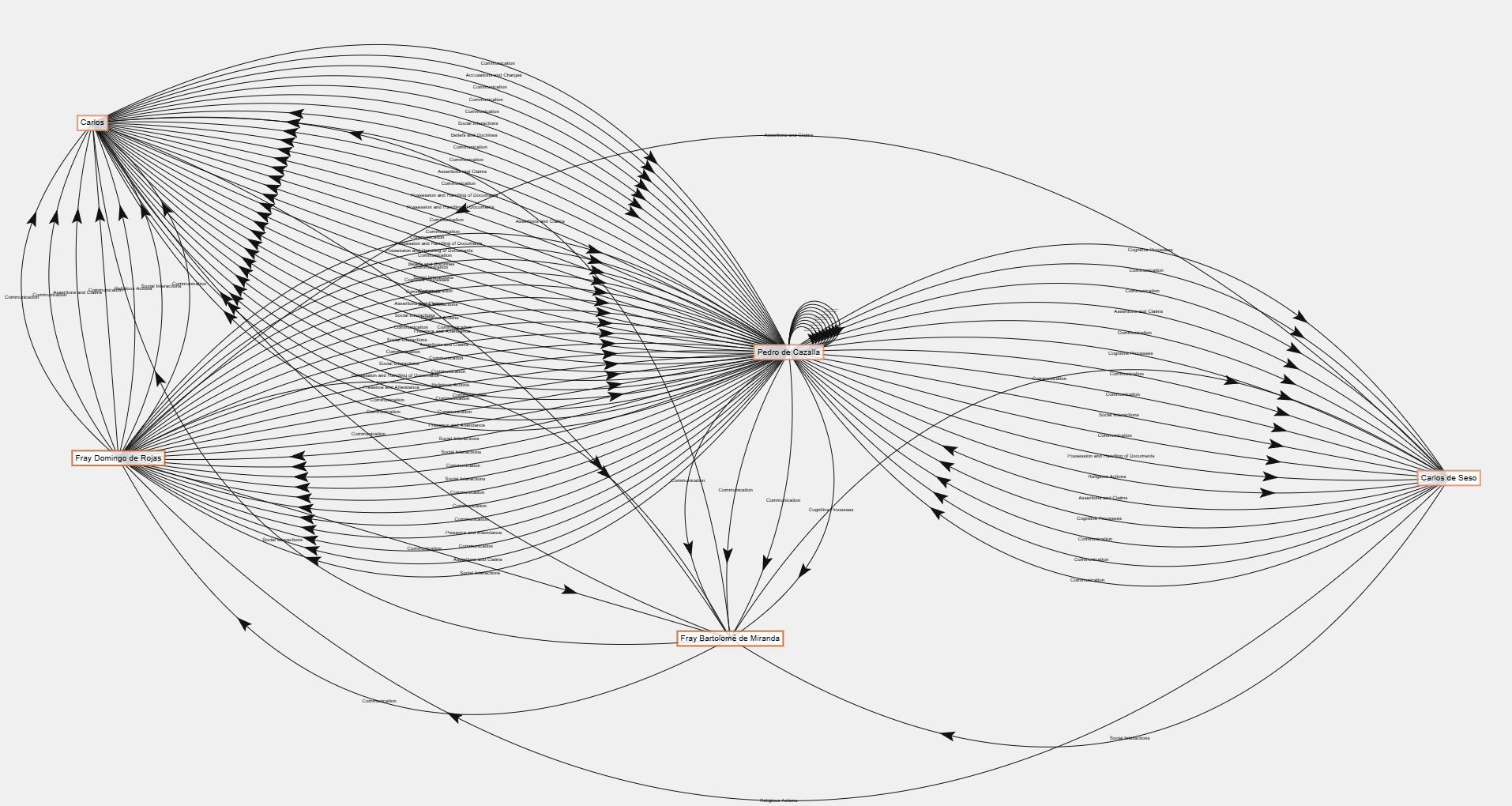} 
\caption{The graph generated by the second query identifies individuals who interacted with Fray Bartolomé de Miranda and quantifies their level  of communication. Note:  the distinction of data labels in this illustration is not essential for understanding.} 
\label{fig:query2}
\end{figure}

\begin{figure}
\centering 
\includegraphics[width=\linewidth]{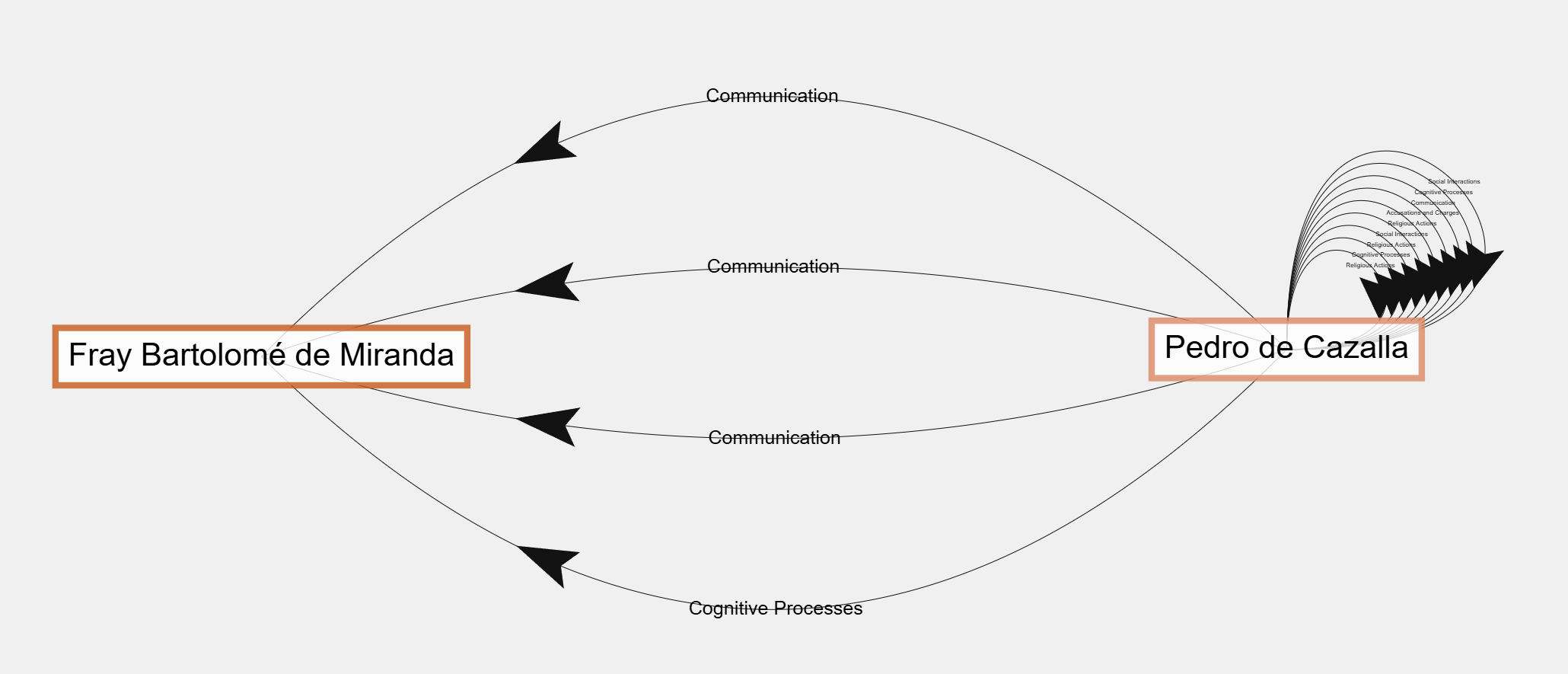} 
\caption{The graph generated by the third query displays all interactions  between Fray Bartolomé de Miranda and Pedro de Cazalla across the entire corpus.} 
\label{fig:query3}
\end{figure}

\begin{figure}
\centering 
\includegraphics[width=\linewidth]{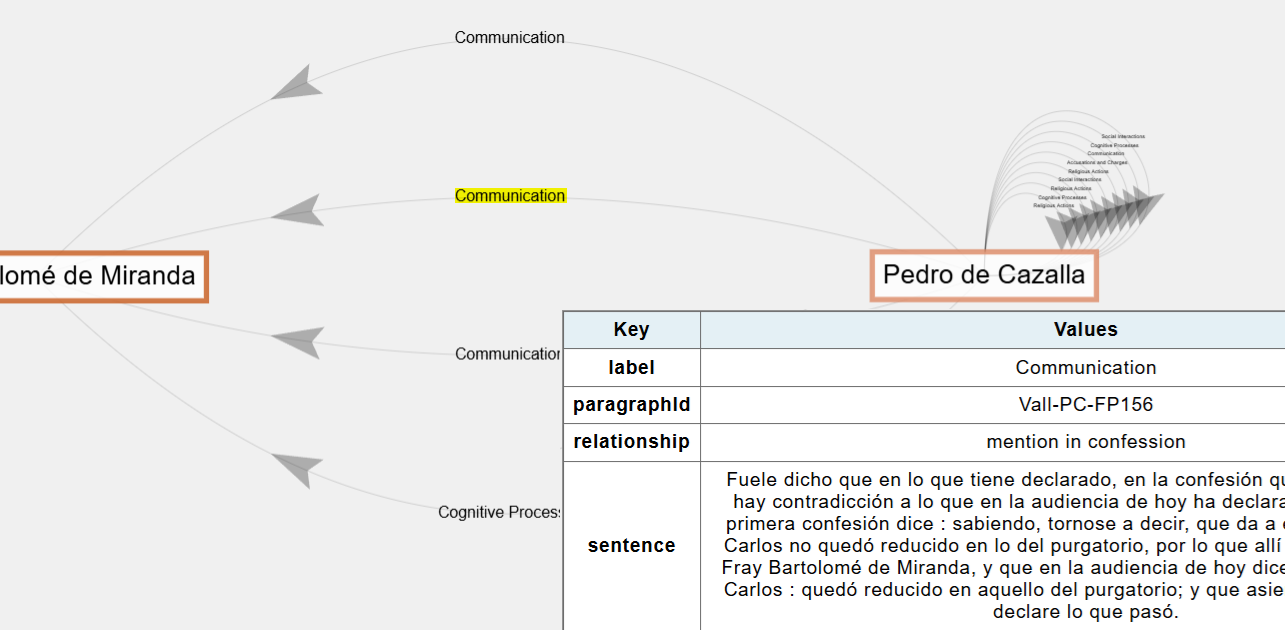} 
\caption{The contextual menu allows users to trace the anchoring of an extracted relationship within the corpus.} 
\label{fig:query4}
\end{figure}

The D4R platform enables historians to explore a large text corpora, which would be impractical to analyse manually, by formulating  natural language queries. Here, we present an example of how a historian applies a scientific method for incremental corpus exploration, refining queries  to gradually uncover a specific pattern.
%
Faced with  a massive dataset, the historian  begins  by displaying recurring and significant historical figures, optionally filtering out those without a religious role (see Figure~\ref{fig:query1}).
Among the key figures appearing in multiple relationships,  Fray Bartolomé de Miranda emerges as a person of interest. A follow-up query  isolates the characters connected to him and examines the nature of these relationships (see Figure~\ref{fig:query2}).
The results indicate that Fray Bartolomé de Miranda  had a direct connection with Pedro de Cazalla, the main defendant in the  Inquisition trial under study. To investigate further, the  historian narrows the search  to focus  exclusively on  these two individuals (see Figure~\ref{fig:query3}).
Each edge in the subgraph represents a specific interaction between Miranda et de Cazalla. At this stage, the historian can  use the contextual menu to access the original texts (see Figure~\ref{fig:query4}). In this final phase, they examine the original text excerpts, either validating their initial hypothesis or refining their interpretation  of the relationship  within the corpus.

This use case illustrates both the historian’s working methodology, as they gradually construct a representation of their research question, and the role of the D4R platform, which enables them to analyse the entire corpus efficiently, regardless of its size.
In conclusion, it is important to emphasize that artificial intelligence does not replace the expertise of historians and doesn't make decisions on their behalf. Instead, it enhances their  capabilities by processing and organising vast amounts of information, providing an efficient interface that seamlessly integrates  into the research workflow.

\section*{Acknowledgement }
This work was supported by the French Agence Nationale pour la Recherche  [grant number ANR-21-CE38-0011]. More information about the project can be found at~\url{https://anr.fr/Projet-ANR-21-CE38-0011}

\bibliographystyle{unsrt}  
\bibliography{main}

\end{document}